\documentclass[article,twocolumn,showpacs,preprintnumbers,amsmath,amssymb]{revtex4}

\usepackage{amsmath,amssymb}
\usepackage{graphicx}
\usepackage{dcolumn}
\usepackage{bm}
\newcommand{\mr}{\mathrm}

\begin{document}
\title{Compensation of Field-Induced Frequency Shifts in Ramsey Spectroscopy of Optical Clock Transitions}
\author{A. V. Taichenachev and V. I. Yudin}
\affiliation{Institute of Laser Physics SB RAS, Novosibirsk
630090, Russia} \affiliation{Novosibirsk State University,
Novosibirsk 630090, Russia} \affiliation{Novosibirsk State
Technical University, Novosibirsk 630092, Russia}
\author{C. W. Oates, Z. W. Barber\footnote{Present address: Spectrum Lab,
Montana State University, Bozeman MT 59717 {USA}}, N. D. Lemke,
and A. D. Ludlow}\affiliation{National Institute of Standards and Technology, Boulder, CO 80305, USA}
\author{U. Sterr, Ch. Lisdat, and F. Riehle}\affiliation{
Physikalisch-Technische Bundesanstalt, Bundesallee 100, 38116 Braunschweig, Germany}
\date{\today}

\begin{abstract}
We have extended Ramsey spectroscopy by stepping the probe frequency during the two Ramsey excitation pulses to compensate frequency shifts induced by the excitation itself. This makes precision Ramsey spectroscopy applicable even for transitions that have Stark and Zeeman shifts comparable to the spectroscopic resolution.
The method enables a new way to evaluate and compensate key frequency shifts, which benefits in particular, optical clocks based on magnetic field-induced, spectroscopy, two-photon transitions, or heavily forbidden transitions.
\end{abstract}
\maketitle

The last few years have been marked by considerable advancements in high-resolution spectroscopy and fundamental laser metrology. Recently optical atomic clocks based on either trapped ions \cite{ros08} or a large number of neutral atoms confined to an optical lattice \cite{kat03, tak05, lud08} have attained uncertainties below those of the best microwave clocks. These efforts could ultimately lead to clocks with fractional frequency uncertainties well below one part in 10$^{17}$. As we push these clocks toward higher performance levels, we continually look for
new spectroscopic methods and approaches that can improve existing atomic clocks and/or enable new types of clocks.

Spectroscopy of ultranarrow, highly forbidden optical transitions like magnetically induced and multiphoton transitions of the $^1S_0$ - $^3P_0$ transition \cite{tai06} in divalent atoms or ions with zero nuclear spin, two photon transitions \cite{bad06,her09,fis04} or octupole transitions \cite{hos09} with nanohertz natural linewidth suffers from excitation related level shifts that limit their applicability for optical clocks.

Typically Rabi excitation of the transition is used due to the simple spectra and the reduced probe light induced ac-Stark shifts. To evaluate and correct the shifts then requires systematic changes in the excitation parameters and the extrapolation to zero, which is difficult, as to cleanly separate various influences.
With two-pulse (Ramsey) excitation \cite{ram50} shifts can be evaluated by changing the dark time between the pulses, which is much more controllable.

However typically in Ramsey spectroscopy excitation related shifts are large and can even distort the signal. In special cases like Electromagnetically Induced Transparency (EIT) \cite{zan06a} or when two shifts of opposite signs occur \cite{yud09_pc} methods have been proposed to cancel these shifts.

In this work, we propose a universal method for the compensation of excitation related shifts that is applicable to any narrow transition. In this method we step the frequency of the probe light during the excitation pulses to cancel excitation-induced shifts. This makes the Ramsey method viable for a wider range of precision optical spectroscopy experiments.

The technique of Ramsey spectroscopy gives precision metrologists an extra parameter, the Ramsey dark time, that can be varied in order to evaluate key frequency shifts \cite{deg05,zan06a}. This allows us to evaluate the effect of a given parameter (e.g., probe light) without varying the physical parameter itself, which can sometimes be experimentally problematic at high levels of precision. Here we quantify the shift(s) by changing the size of
the frequency step of the laser for different Ramsey dark times (duration and frequency are easily controlled experimental quantities), until we find the cancellation point. In fact, we can null multiple shifts simultaneously, thereby reducing the number of parameters that need to be evaluated. Additionally, compensation of the shifts during excitation is critical for variants on the standard Ramsey pulse sequence that directly
reduce these key shifts (and their associated uncertainties) rather than just compensate them; these will be described in an upcoming publication. Finally, Ramsey excitation can lead to improved stability when the measurement duty cycle is less than 100\% \cite{dic87,que03}. Thus, we anticipate that shift-compensated Ramsey spectroscopy could accelerate progress for some of the most promising technologies, as well as open up the possibility of using atomic systems such as two-photon transitions, for which field-induced frequency shifts have
presented a considerable barrier \cite{hal89}.

\begin{figure}[t]
	\centerline{\scalebox{0.45}{
		\includegraphics{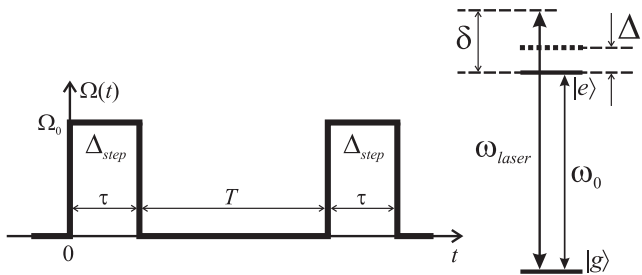}}}
	\caption{Fig.~1. Illustration of the two pulse Ramsey sequence. $\Omega$ is the excitation Rabi frequency and is shown as a function of time. During the pulses we step the laser frequency by $\Delta_\mr{step}$ in an attempt to match shift of the transition frequency ($\Delta$) due to effects of the excitation fields. During the dark period between pulses ({\it{T}}), the atomic transition and laser field frequencies are unperturbed and evolve at their unshifted values. Also shown is a two level atom with bare levels split by $\omega_0$ and an induced shift $\Delta$.} \label{Ramfig}
\end{figure}

Under usual conditions for Ramsey spectroscopy we assume fixed frequencies for the excitation source and atom resonance throughout the excitation and dark periods (see Fig.~\ref{Ramfig}). The net perturbation caused by small frequency shifts induced by the excitation (e. g., an ac Stark shift due to probe light) can be estimated from the product of the size of shift $\Delta$ and the ratio of excitation time to total Ramsey time 2$\tau$/($T+2\tau$). Here let us consider instead Ramsey spectroscopy for a two-level system when there is a more significant frequency shift induced by the excitation field (which we assume is independent of detuning around the narrow resonance).
That is, we include in the derivation of the Ramsey lineshape a frequency shift $\Delta$ that acts only during the excitation pulses. If the atoms are initially in the lower level $| g \rangle$, then the population of atoms in the excited state $| e \rangle$ after interacting with the two pulses is determined by the following expression:
\begin{eqnarray}\label{n_e}
 n_{\mr{e}}  =  && \frac{\Omega_0^2}{\Omega^2}\times
 \left[\cos\left(\frac{\delta \cdot T}{2}\right)\sin(\Omega\tau) \right. \\
 && \left. -\frac{2(\delta-\Delta)}{\Omega}\,\sin\left(\frac{\delta \cdot T}{2}\right)
\sin^2\left(\frac{\Omega\tau}{2}\right)\right]^2, \nonumber
\end{eqnarray}
where $\delta = \omega-\omega_0$ is the detuning of the probe field frequency from the frequency of the unperturbed transition (i.e. during free evolution between Ramsey pulses), $\Omega_0$ is the Rabi frequency, and $\Omega =
\sqrt{\Omega_0^2+(\delta-\Delta)^2}$ is the generalized Rabi frequency.

The formula (\ref{n_e}) describes typical Ramsey fringes, where the location of the central resonance (as a function of $\delta$) is the key parameter for our discussion. The presence of the additional frequency shift $\Delta$ during the pulses leads to a shift of position of the central resonance maximum $\overline{\delta\omega}_0$ with respect to the frequency of the unperturbed transition $\omega_0$. To find an analytical expression for the dependence of $\overline{\delta\omega}_0$ on $\Delta$, we determine the position of the maximum of the central Ramsey fringe. We present the signal $n_{\mr{e}}$ as a Taylor series with respect to the dimension-less parameter $| \delta \cdot T | \ll 1 $ involving the detuning $\delta$:
\begin{equation}
n_{\mr{e}}=a^{(0)}+a^{(1)}(\delta \cdot T)+a^{(2)}(\delta \cdot T)^2+...\,,
\label{eq:n_e_d}
\end{equation}
where the coefficients $a^{(i)}$ can be expanded in the powers of $\Delta/\Omega_0$ in the following way:
\begin{eqnarray}
a^{(0)}&=&{\cal A}_{0}^{(0)}+{\cal A}_{2}^{(0)}\left(\frac{\Delta}{\Omega_0}\right)^2+{\cal A}_{4}^{(0)}\left(\frac{\Delta}{\Omega_0}\right)^4+...\nonumber\\
a^{(1)}&=&{\cal A}_{1}^{(1)}\left(\frac{\Delta}{\Omega_0}\right)+{\cal A}_{3}^{(1)}\left(\frac{\Delta}{\Omega_0}\right)^3+...\nonumber\\
a^{(2)}&=&{\cal A}_{0}^{(2)}+{\cal A}_{2}^{(2)}\left(\frac{\Delta}{\Omega_0}\right)^2+{\cal A}_{4}^{(2)}\left(\frac{\Delta}{\Omega_0}\right)^4+...
\label{eq:Aj}
\end{eqnarray}
The occurrence of terms with odd or even powers only is the direct consequence of the symmetry of Eq.~(\ref{n_e}) that does not change under the simultaneous substitutions $\delta \to -\delta$ and $\Delta \to -\Delta$.

For $|\Delta/\Omega_0| \ll 1$ we find from Eq.~(\ref{eq:n_e_d}) the leading dependence of $\overline{\delta\omega}_0$ on the coefficients $a^{(i)}$:
\begin{equation}
\overline{\delta\omega}_0\approx-\frac{1}{T} \frac{a^{(1)}}{2a^{(2)}} \, .
\label{eq:d0}
\end{equation}
Under the condition $|\Delta/\Omega_0| \ll 1$ this shift $\overline{\delta\omega}_0$ has the following form:
\begin{eqnarray}\label{shift1}
&&\overline{\delta\omega}_0\approx\Delta\times\\
&&\frac{\Omega_0 T- \Omega_0(T+2\tau)\cos(\Omega_0\tau)+2\sin(\Omega_0\tau)}{2\Omega_0 T - 2\Omega_0(T+\tau)\cos(\Omega_0\tau)+(2+\Omega_0^2T^2/2)\sin(\Omega_0\tau
)}\,.\nonumber
\end{eqnarray}
Also, if $|\Delta/\Omega_0|\ll 1$, the amplitude of the central resonance is maximal ($\approx\!1$) for $\Omega_0\tau = (2l+1)\pi/2$ (where $l$~=~0,~1,~2,...). For the usually chosen case of $l=0$, the shift (\ref{shift1}) can be written as
\begin{equation}\label{sh_res}
\overline{\delta\omega}_0\approx
\frac{\Delta}{1+(\Omega_0 T/2)}=\frac{\Delta}{1+(\pi/4)(T/\tau)}\,,
\end{equation}
i.e., for $0\leq T <\infty$ the shift $\overline{\delta\omega}_0$ monotonically decreases from $\Delta$ to 0. For $(T/\tau)\gg 1$ one can rewrite (\ref{sh_res}) by the following expression:
\begin{equation}\label{sh_res2}
\overline{\delta\omega}_0\approx \frac{4}{\pi}\,\frac{\tau}{T}\,\Delta=\frac{2}{T}\,\frac{\Delta}{\Omega_0}\,.
\end{equation}
In order for Eqs.\,(\ref{shift1}) to (\ref{sh_res2}) to be valid we need to have $|\Delta/\Omega_0| \ll 1$, i.e., the Fourier width larger than the shift. Otherwise (i.e., for $|\Delta/\Omega_0|>1$) the line is significantly shifted and the central resonance amplitude can be greatly reduced, making the Ramsey method unattractive.  In cases such as two-photon spectroscopy and Magnetically Induced Spectroscopy (MIS), the atomic resonances are shifted by the excitation fields by an amount that can be roughly comparable to $\Omega_0$. Thus, it is
necessary to have an additional shift that counteracts this induced shift, so that during the excitation pulses the net shift is much less than $\Omega_0$. There are several ways we can generate this cancellation field. The most general and straightforward way is to step the frequency of the laser field by $\Delta$ (i.e., $\omega'_{laser} = \omega_{laser} + \Delta$) during the excitation. Then, during the excitation pulses the atom resonance and laser frequency are shifted by equal amounts and the Ramsey spectroscopy proceeds efficiently. During the dark time,
$T$, the laser frequency is returned to its unshifted value (in a phase coherent way) as the atomic resonance is also unshifted by excitation field(s) during this period. Even though the atoms do not see the light during the dark time, it is critical that the laser light has the appropriate phase for the second pulse. Thus, another way to look at this variant of Ramsey spectroscopy is that here an atomic interferometer is used to determine when (i.e., for what value of the laser frequency shift) the atomic phase evolution is the same during the interrogation and dark times. Conversely with Rabi excitation, there is only an excitation pulse, thereby preventing this comparison of phase evolutions. Experimentally, the phase coherent shift of the laser light can readily be accomplished with an acousto-optic modulator controlled by a direct digital synthesizer. Alternatively, the frequency can be fixed throughout the spectroscopy (at the shifted value) and the phase could be appropriately stepped before the second Ramsey pulse.

An alternative way to generate the needed cancellation is available for types of spectroscopy that use multiple excitation fields that generate shifts with opposite signs. This is the case for MIS, which uses a light field and magnetic field in tandem to excite atoms, with Yb atoms confined in a standing wave optical lattice \cite{bar06}. The second-order Zeeman and ac Stark shifts associated with the fields used to excite the Yb clock transition
can be set to be equal and opposite, so the net shift is zero. Of course this requires that the magnetic field be turned on only during excitation, which could be experimentally challenging for short Ramsey pulses. It would be easier to step the frequency instead, but as we will see there could be some advantages to folding both shifts into the excitation pulses. For MIS applied to Sr lattice clocks, however, the Zeeman and Stark shifts have the
same signs, so the frequency stepping method seems appropriate. A third method would be to use an additional, fairly far-detuned laser to produce a Stark shift with suitable sign during the excitation pulses. This technique is quite general and could be used instead of frequency stepping, although it is more complicated experimentally. Note that far-detuned lasers have been used previously to generate advantageous Stark shifts in microwave
clock \cite{kap02} and dipole trap loading \cite{gri06} experiments.

\begin{figure}[t]
\centerline{\scalebox{0.5}{
		\includegraphics{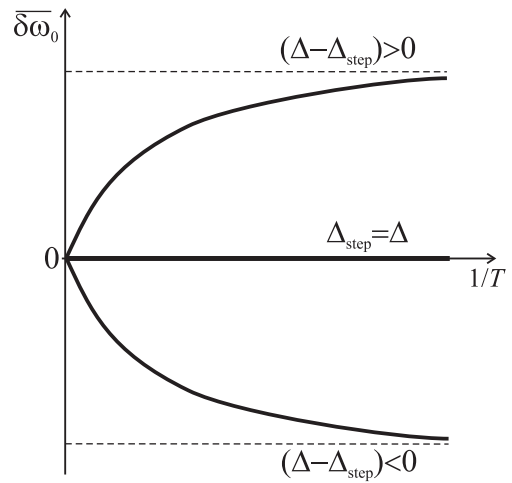}}}
	\caption{Fig.~2. Theoretical shifts of the Ramsey maximum as a function of the inverse of the Ramsey dark time, $T$, for different values of the frequency step,
$\Delta_\mr{step}$. For $\Delta_\mr{step}=\Delta$, the line has zero
slope and the shift is cancelled.}\label{Darktimelines}
\end{figure}

With the condition $|\Delta/\Omega_0| \ll 1 $ fulfilled by one of the above methods, we can take advantage of the additional degree of freedom (i.e., the dark time) provided by the Ramsey method in several different ways. The simplest is to set the values of the excitation fields such that they approximately cancel. Assume for
example that we use the frequency stepping method. Then a series of measurements of the frequency of the Ramsey maximum vs. the Ramsey dark time relative to a sufficiently stable flywheel oscillator would determine the residual offset in the cancellation. Since the shift is fixed during the excitation pulses, changing the dark time will change only the fraction of time the shift is present during the whole spectroscopic period (duration of two pulses plus that of the dark time). We thus expect a linear relationship between the residual shift and $1/T$,
extrapolating to zero for infinite Ramsey time (recall Eq.~(\ref{sh_res2})). By making several measurements at different parameter values, we can quickly zoom in on the cancellation values (see Fig.~\ref{Darktimelines}).

\begin{figure}[t]
\centerline{\scalebox{0.5}
{\includegraphics[width=16cm]{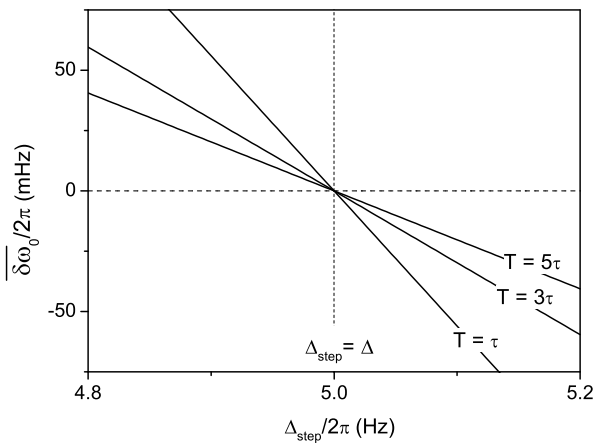}}}
\caption{Fig.~3. Shift of the Ramsey maximum $\overline{\delta\omega}_0$ as a function of frequency step $\Delta_\mr{step}$, for several different values of the dark time $T$. Conditions similar to an operating ytterbium lattice clock \cite{pol08} are assumed: Rabi frequency $\Omega/2\pi = 5$~Hz, pulse length $\tau = 50~$ms and probe laser ac-Stark shift of $\Delta/2\pi \approx 5~$Hz. The quadratic Zeeman shift is neglected. The curves intersect at zero shift, where $\Delta_\mr{step}=\Delta$.}
\label{shifts}
\end{figure}

Alternatively, a series of measurements of the frequency of the Ramsey maximum  vs. different step values for two or three different Ramsey conditions would yield a series of lines that would intersect at the cancellation step frequency (see Fig.~\ref{shifts}), $\Delta_\mr{step}=\Delta$. Then with the system set to the ``magic'' step frequency and optimal pulse duration ($\Omega_0\tau = \pi/2$), locking the laser frequency to the Ramsey maximum would yield a probe field frequency that is stabilized on the unperturbed resonance frequency $\omega_0$.

The main advantage of this technique is a new capability for investigating and suppressing frequency shifts that occur during excitation. With the Rabi spectroscopy method, the evaluation of these shifts usually requires a series of measurements of the resonance frequency over a range of intensities of the excitation fields, whose changes could have unwanted side effects on the experiment. (For example, magnetic field shifts in MIS need to be
evaluated over a large range of magnetic field values that need to be tuned and controlled to a high degree). The shift under examination then needs to be accounted for (usually after the fact). With this new method we measure {\it {and null}} the net effect of the field-induced shifts. Because this technique does not rely on an extrapolation to zero based on a series of measurements that yield a non-zero slope (as occurs with with
Rabi-based measurements of the ac Stark and the Zeeman effect), we have more flexibility. For example, if drifts of this value over long periods (e.g., during a 24-hour clock comparison) are a concern, occasional interspersed measurements of the frequency at different Ramsey dark times could be used to check for shifts. We
could even servo the step frequency to keep this shift nulled in almost real time. Additionally, in certain cases we can null multiple shifts (such as the laser- and magnetic-induced shifts in MIS) simultaneously.

What is not immediately evident from this discussion is whether this method significantly reduces the resultant uncertainties for the relevant parameters.  If the uncertainty associated with an excitation field depends on our ability to evaluate the residual shifts (rather than on our degree of control over the shift), the choice between Rabi and Ramsey may depend on the idiosyncracies of a given experiment. For the case of the ac Stark shift of the
probe light, the Ramsey method still suffers from requiring larger fields (and thus larger Stark shifts), but they can be evaluated by just turning a knob ($\Delta_\mr{step}$) that controls the probe field frequency to achieve a null condition that requires no extrapolation. To reduce uncertainties on this null value, we can generate a large leverage factor by using short Ramsey interaction times (e.g., in Eq.~(\ref{sh_res}), the shift becomes smaller with
smaller $\tau /T$). But this advantage is partially offset by the fact that to achieve a short $\tau$ requires larger fields, which could in turn lead to larger uncertainties due to limitations in experimental control. Ultimately, the preferred technique will be determined by measurements in the laboratory, with the possibility
of redundancy offered by the two methods always available.

This technique could also be applied to the case of the two-photon spectroscopy when the probe field frequency equals the half of the clock transition frequency \cite{vas70}. For example, the dipole-forbidden transitions $^1$S$_0 \to ^1$S$_0$ or $^1$S$_0 \to ^1$D$_2$ can be used. Here it is assumed that the higher levels
$^1$S$_0$ and $^1$D$_2$ are long-lived (as, for example, $^1$S$_0$ for He and $^1$D$_2$ for Ca and Ba). Two-photon transitions have fallen somewhat out of favor as candidates for state-of-the-art frequency standards, in part because they have Stark shifts that are comparable to the two-photon Rabi frequency \cite{vas70}. However, with the advent of hertz-wide laser sources, the associated shifts no longer appear so formidable, so these transitions  may again be viable. The shift-compensated Ramsey spectroscopy proposed here is well suited for handling these
shifts and thereby would enhance new possibilities for optical atomic clocks.

In conclusion, we have presented a variant on the Ramsey method that enables Ramsey spectroscopy to be used with spectroscopic systems that have large excitation induced shifts. This method relies on the use of an additional cancellation shift during the excitation pulse and can be easily implemented. Moreover, this technique has considerable generality and could be used with single-photon, two-photon, or magnetic field-induced spectroscopy.
Because of its flexibility, this method could enable new clock transitions as well as benefiting some existing systems \cite{hos09, pol08, bai07}, since it makes accessible the usual features of Ramsey spectroscopy and also gives a experimentalist another tool with which to address the problem of excitation-induced shifts. The advantage of nulling one or more shifts simultaneously, rather than just evaluating them, may enable more efficient evaluation and reduction of their associated uncertainties. As the evaluation of such uncertainties is a
critical part of precision measurements such as those associated with atomic clocks, this new technique could be useful in a variety of experiments.

The authors thank L. Donley and R. Fox for their careful reading of the manuscript. A.V.T. and V.I.Yu. were supported by RFBR (07-02-01230, 07-02-01028, 08-02-01108), and programs of RAS. V.I.Yu., U.S., Ch.L. and F.R. gratefully acknowledge support by the Centre for Quantum Engineering and Space-Time Research (QUEST).

A.V.T. and V.I.Yu. e-mail address: llf@laser.nsc.ru


\begin{thebibliography}{25}
\bibitem{yud09_pc}
E.g. in a $^{174}$Yb optical lattice clock, the ac-{S}tark shift of the
  excitation laser and the quadratic {Z}eeman shift have opposite signs (see in \cite{tai06}).
\end{thebibliography}
\end{document}